# Galactic origin of ultrahigh energy cosmic rays

A.A. Mikhailov[*]

Yu.G. Shafer Institute of Cosmophysical Research and Aeronomy, 31 Lenin Ave., 677980 Yakutsk, Russia

Abstract

The arrival directions of ultrahigh energy extensive air showers (EAS) by Yakutsk, AGASA, P. Auger array data are analyzed. For the first time, the maps of equal exposition of celestial sphere for the distribution of particles by AGASA and P. Auger arrays data have been constructed. The large-scale anisotropy of cosmic particles at $E>4.10^{19}$ eV by Yakutsk, AGASA and P. Auger array data has been detected. The problem of cosmic particle origin is discussed.

## 1. INTRODUCTION

Until now there is an opinion that cosmic rays with energy $E>4.10^{19}$ eV are isotropic [1,2]. Here data of EAS Yakutsk, AGASA, P. Auger arrays in terms of their exposition of the celestial sphere to arrays are analyzed.

Shower cores of the Yakutsk EAS (extensive air shower) lie inside the array perimeter and the accuracy of arrival directions angle determination is ~3°. The particle energy has been determined with the accuracy ~30%. Three showers have energy $E.10^{20}$ eV. The particle energy by data AGASA is determined with the accuracy ~25%, the solid angle - ≤1.6° [3]. The particle energy by data P. Auger is determined with the accuracy ~22%, the solid angle - ≤1° [4].

## 2. A LARGE-SCALE ANISOTROPY

Fig.1 presents the distribution of 34 particles by Yakutsk array data with the energy $E>4.10^{19}$ eV on the map of equal exposition of the celestial sphere for the period of 1974-2007 (the method to construct this map is based on the estimation of the expected number of showers in the celestial sphere [5,6] and etc.). On the map of equal exposition the equal number of particles from the equal parts of sphere is expected.

In Fig.1 the most concentration of particles is observed from the side of Input of Local Arm Orion of Galaxy at the galactic latitude $3.3°<b<29.7°$ and longitude $60.1°<l<116.8°$ (this region is noted by dash quadrangles). In this coordinates there are 9 particles. We divided the whole period of observation into 2 periods, equal to the exposition of observation: 1974 - 1985 and 1986 - 2007. For the first period 4 particles (circles) have been registered, the probability of chance to find 4 of 34 particles in above-mentioned coordinates using the method [6] is P~0.1.

We take the first period as a sample and for second period 5 particles (triangles) were registered, the probability of chance to find of 34 particles in the coordinates above-mentioned is P~0.03.

For the distribution of particles with energy $E>4.10^{19}$ eV by AGASA array data [3] we've constructed the map of equal exposition of the celestial sphere according to [6]. As it is seen from this map, almost a half of events (25 particles of 58) are within coordinates $11.2°<b<69.3°$ and $38.9°<l<154°$ towards a side of Input of the Local Arm. Also we have divided the whole period of observation into 2 periods, equal to the exposition of observation: 1984 - 1994 and 1995 - 2001.

For the first period 13 particles (circles) were registered, the probability of chance to find 13 of 58 particles in above-mentioned coordinates is P~$10^{-3}$. We take the first period as a sample and for second period 12 particles (triangles) were registered, the probability of chance to find 12 of 58 particles in the above-mentioned coordinates is P~$10^{-3}$.

Thus, the statistically significant particle flux in the case of the AGASA array is observed from the side of Input of Local Arm as well as data of the Yakutsk EAS array.

For the distribution of particles with the energy E>$5.7.10^{19}$ eV of the P. Auger array [4] the map of equal exposition of the celestial sphere have been constructed. As is seen from this map, 7 particles of 27 are within of coordinates towards the side of Output of the Local Arm near a galactic plane: $1.7° \leq b \leq 19.2°$ and $-52.4° \leq l \leq -34.4°$. Also we have divided the whole period of observation into 2 periods, equal to the exposition of observation: 2004 - 2005 and 2006 - 2007. For first period 3 particles (circles) was registered the probability of chance to find 3 of 27 particles in above-mentioned coordinates is P~$10^{-3}$. We take the first period as a sample and for the second period 4 particles (triangles) were registered the probability of chance to find 4 of 27 particles in the above-mentioned coordinates is P~$3.10^{-5}$.

As follows from the distribution of particles on the celestial sphere by the data of the Yakutsk, AGASA, P. Auger extensive air shower arrays (Fig.1-3), there is a large-scale anisotropy in arrival direction of particles connected with the a galactic plane and Local Arm Orion of the Galaxy.

Earlier we found that if sources of particles are distributed uniformly in the Galaxy disc, then protons with the energy E~$10^{18}$ eV mainly move along the magnetic field lines of the Galaxy Arms [7]. Therefore, one can suggest that the observed particles with the energy E>$4.10^{19}$ eV from the side of the Input and Output of Local Arm Orion of Galaxy along the magnetic field lines have a rigidity R~$10^{18}$ eV and they are a charged heavy particles. The similar conclusion about the composition of cosmic rays was made by us on the basis analysis of the distribution of particles in the celestial sphere and the muon content of extensive air showers at energy E=$10^{19}$-$10^{20}$ eV by the Yakutsk EAS array data [8,9]. Also independently from us our colleagues using another method of muon data of showers analysis have made a conclusion [10] that the iron nuclei portion at energy E>$2.10^{19}$ eV can be from ~29% up to 68% of the total number of particles. The last data of the P. Auger array indicate that the particles at energy E>$4.10^{19}$ eV are not only protons, the average mass of particles A is equal lnA=$2.6 \pm 0.6$ [11].

Thus, a large-scale anisotropy of particles at E>$4.10^{19}$ eV from the side of Local Arm Orion of Galaxy has been found.

## SIGN OF THE GALACTIC ORIGIN

In the paper [12] a galactic model of cosmic ray origin of ultrahigh energy has been considered. It is supposed, that sources of particles are uniformly distributed over the Galaxy disc, the regular magnetic field of the Galaxy has mainly azimuth directions. If a magnetic field doesn't change its direction above and below a galactic plane (near the Sun a magnetic field is directed in galactic longitudes l ~ 90°, then as a result of influence of large-scale regular magnetic field, characteristic trajectories of particles are possible. (Fig. 4).

In the case of uniform distribution of cosmic ray sources in the Galaxy disc an expected particle fluxes by separate directions of the celestial sphere will be proportional to lengths of particles trajectories in this direction. As follows from Fig.4 from the center of the Galaxy with positive latitudes and from the anticenter from negative latitudes the increased fluxes of particles are expected.

Therefore, from the side of Galaxy center (longitudes - 90°<l<90°) the ratio of the number of particles above the galactic plane to the number of particles below the plane will be $n_c(b>0°)/n_c(b<0°)>1$ and from the side of anticenter the ratio will be $n_a(b>0°)/n_a(b<0°)<1$. This is a sign of galactic origin of cosmic rays.

We've considered the distribution of particles in galactic latitudes by data the Yakutsk, AGASA, P. Auger for sides: the center and the anticenter of the Galaxy. The particle distribution by data of the northern arrays Yakutsk, AGASA is considered separately (Fig.5) from the data of P. Auger, which is located in the southern hemisphere (Fig.6). The total number of particles is equal to 119.

In Fig.5a,c the distribution of particles in galactic latitudes above/below a galactic plane for the Galaxy center is shown. The number of particles in the subsequent interval of angles have been summarized. The expected number of particles (curve) in the case of isotropy is shown [6]. Also in Fig.5b,d the distribution of particles in galactic latitudes above/below a galactic plane for the Galaxy anticenter is also shown.

According to Fig.5 the increased flux of particles is observed at latitudes $|b|>15°$ from the center above a plane and from the anticenter below a plane of the Galaxy (Fig.5a, d), the decreased flux is observed at $|b|>30°$ from the center at the below a plane and from the anticenter the above a plane of the Galaxy (Fig.5c,b). Practically the same particle distributions in galactic latitudes are observed by data of the P. Auger (Fig.6). The independent data of the southern array confirm the particle distribution in galactic latitude of the northern arrays.

In Fig.7 ratio of the number of particles above galactic plane $n_u$ to the number of particles below the galactic plane $n_b$ in terms of the exposure of the celestial sphere is shown YA, $A=n_u/n_b=n_1(b>0°)S_2/n_2(b<0°)S_1$ at the interval of angles <90° from the galactic plane, where data of arrays is marked - YA = Yakutsk + AGASA, A = P. Auger, $n_1$, $n_2$ are the number of particles, $S_1$ and $S_2$ are the exposure of the celestial sphere to the arrays Yakutsk, AGASA, P. Auger. Also the ratio of the number of particles by 3 arrays above/below the galactic plane R (Yakutsk + AGASA + P. Auger) at the center and the anticenter of the Galaxy in terms of the exposure of the celestial sphere to these arrays is shown.

We've considered these ratios YA, A, R in two cases (Fig.7): the center (a) and the anticenter of the Galaxy (b). The ratio of the numbers of particles YA, A, R from the side of the center/anticenter differs: from the center – $(YA)_c$, $A_c$, $R_c>1$, from the anticenter – $(YA)_a$, $A_a$, $R_a<1$. As we have mentioned above this is a sign of galactic origin of cosmic rays.

In Fig. 7 the ratio of the number of particles $R_c$ is not a maximum value because this value was found at interval of angles <90° from the galactic plane. The maximum value $R_c$ will be between 0- 90° from the galactic plane (see a distribution lengths of trajectories in Fig.4). For example, value $R_c$ at interval of angles <45° (half of interval 90°) from the galactic plane has increased $R_c \sim 1.55$ (instead of 1.4 at interval <90°, Fig.7a) and in a case anticenter $R_a$ it has decreased $R_a \sim 0.75$ (instead of 0.85, Fig.7b).

Note, the probability of chance P to observe particles by data of 3 arrays by the following sides: ≥34 particles (center, b>0°), ≥30 particles (anticenter, b<0°), ≤40 particles (anticenter, b>0°), ≤15 particles (center, b<0°) is $P \sim 10^{-2}$ (at interval of angles <45° - $P \sim 2 \cdot 10^{-3}$). This probability was found by simulation 119 events over the whole celestial sphere in terms exposure of the celestial sphere to arrays [6].

The ratio of the number of particles R from the side of center/anticenter at interval of angles <90° is $R_c/R_a \sim 1.7\pm0.6$ (at the interval of angles <45° - $R_c/R_a \sim 2\pm0.8$). This ratio can be interpreted as follows: from the side of center the density and region of sources along trajectory particles are higher and larger than the density and region of sources from the anticenter of the Galaxy. Such distribution of sources it is a possible only in a galactic origin of particles. Thus ratio $R_c/R_a$ confirms our conclusion about the galactic origin of cosmic rays (see above).

We also note that the distribution of particles in galactic latitude relatively an expected distribution particles in the case of isotropy depends on sides of the galactic plane (above/below and center/anticenter, Fig. 5-7). It means that an influence of the magnetic field of he Galaxy on the movement of particles of ultrahigh energy is rather strong.

## 4. CONCLUSION

Most likely the main part of particles with $E>4.10^{19}$ eV has a galactic origin.

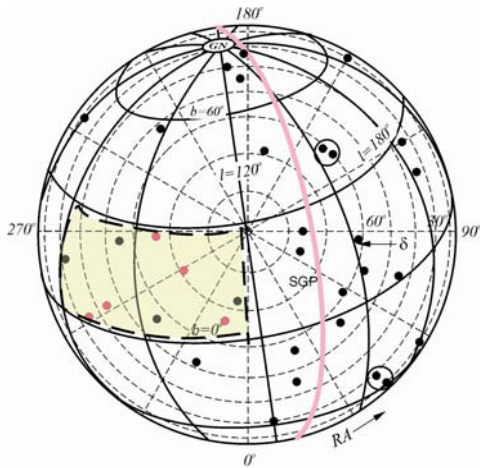

Fig.1. On the map of equal exposition particles with energy $E>4\cdot10^{19}$ eV are shown by Yakutsk EAS array data. SGP - Super Galactic Plane. Dashed quadrangles on the left – a considered region. δ - declination, RA - right ascension, b, l - a galactic latitude and longitude. Large circles -clusters. Inside quadrangles: ● - 1974-1985, ● - 1986-2007.

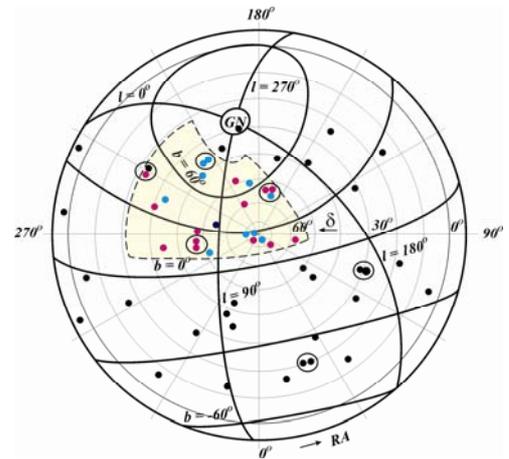

Fig.2. The same as in Fig.1 for the AGASA array data. Dashed quadrangles on the upper - a considered region of a celestial sphere. Inside quadrangles: ● -1984-1994, ● - 1995-2001.

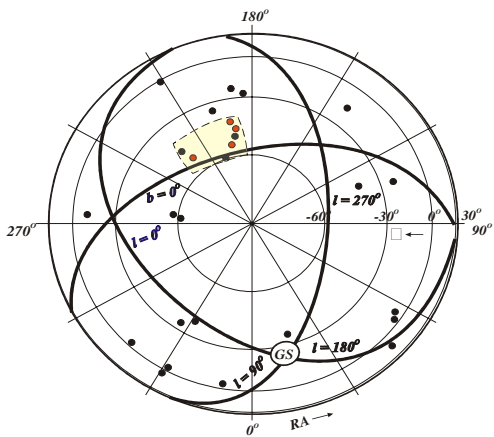

Fig.3. The same as in Fig.1 for the P. Auger array data. Dashed quadrangles - a considered region of a celestial sphere. Inside quadrangles: ● - 2004-2005, ●- 2006-2007.

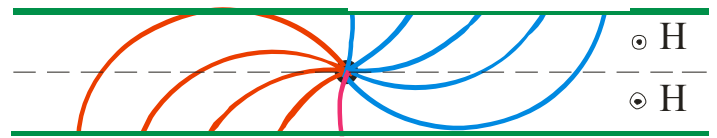

Fig.4. Trajectories of particles in a magnetic field H of a disc, ⊙ - a magnetic field.

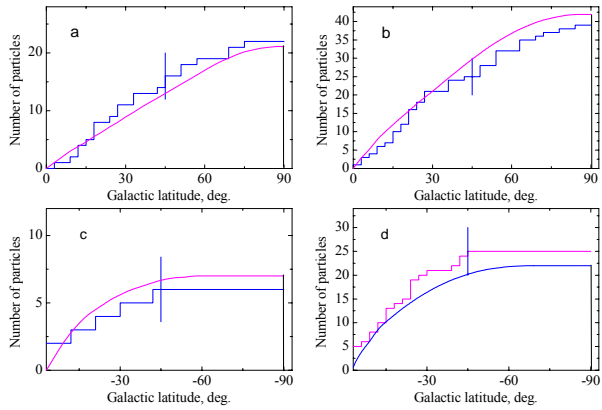
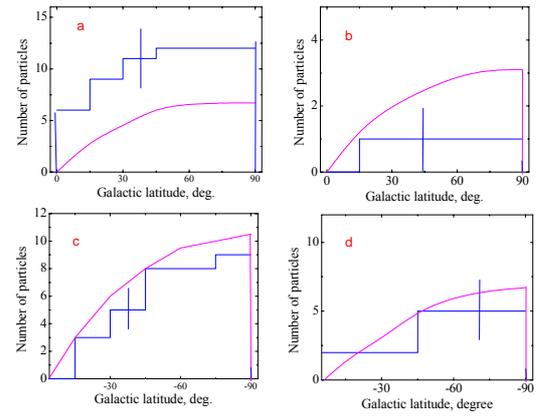

Fig.5. Distribution particles (Yakutsk+AGASA) versus a galactic latitude in sides: a, c - center, b, d - anticenter of Galaxy. The curve line is the expected number of particles in the case of isotropy.

Fig.6. Distribution of the observed and the expected number particles by P. Auger data the same directions as Fig.5.

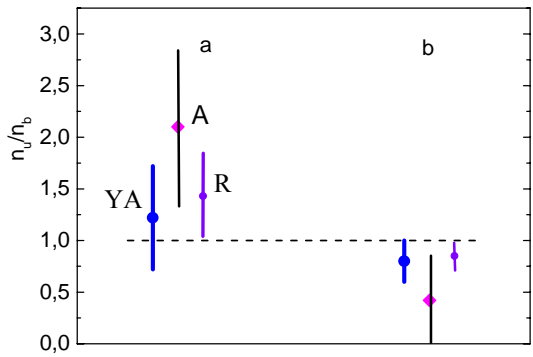

Fig.7. Ratio number of particles above/below of a galactic plane. ● - Yakutsk+AGASA, (YA), ♦ - P. Auger, (A) , ● - Yakutsk+AGASA+P. Auger, (R), a - center, b - anticenter.